%% file: main.tex
\documentclass[sigconf, nonacm]{template/aamas} 

\usepackage{balance}
\usepackage{float}

\usepackage{todonotes}
\usepackage{graphicx}
\usepackage{algorithm}
\usepackage[noend]{algpseudocode}
\usepackage{enumitem}

\input{common/theorem}
\input{common/glossary}

\input{common/code_reference_statement}
\input{common/common_definitions}

\title[Enhancing the Action Dependency Graph Framework]{Streamlining the Action Dependency Graph Framework: Two Key Enhancements}

\author{Joachim Dunkel}
\affiliation{
  \institution{Graz University of Technologies}
  \city{Graz}
  \country{Austria}}

\input{src/abstract.tex}

\keywords{MAPF, ADG, Runtime Optimization, Multi-Agent Systems, Plan Execution}

\newcommand{\BibTeX}{\rm B\kern-.05em{\sc i\kern-.025em b}\kern-.08em\TeX}

\begin{document}

\pagestyle{fancy}
\fancyhead{}

\maketitle 

%%%%%%%%%%%%%%%%%%%%%%%%%%%%%%%%%%%%%%%%%%%%%%%%%%%%%%%%%%%%%%%%%%%%%%%%

\input{src/introduction}

\input{src/problem_description}

\input{src/approach}

\input{src/experimental_evaluation}

\input{src/conclusion}

\bibliographystyle{template/ACM-Reference-Format} 
\bibliography{citations/manual_citations, citations/zotero_citations, sample}

%%%%%%%%%%%%%%%%%%%%%%%%%%%%%%%%%%%%%%%%%%%%%%%%%%%%%%%%%%%%%%%%%%%%%%%%

\end{document}

%% file: common/theorem.tex
\usepackage{amsthm} 

\newtheorem{theorem}{Theorem}      
\newtheorem{definition}{Definition}
\newtheorem{proposition}{Proposition} 
\newtheorem{corollary}{Corollary}

%% file: common/glossary.tex
\usepackage[acronym]{glossaries}

\newcommand{\glslong}[1]{\glsreset{#1}\gls{#1}}

\newacronym{mapf}{MAPF}{Multi-Agent Path Finding}
\newacronym{mapd}{MAPD}{Multi-Agent Pickup and Delivery}
\newacronym{adg}{ADG}{Action Dependency Graph}
\newacronym{seAdg}{SE-ADG}{Spatially Exclusive Action Dependency Graph}
\newacronym{sAdg}{SADG}{Switchable Action Dependency Graph}
\newacronym{rhcre}{RHCRE}{Rolling Horizon Collision-Resolution and Execution}
\newacronym{rhcr}{RHCR}{Rolling Horizon Collision Resolution}
\newacronym{mapfPost}{MAPF-POST}{MAPF-POST} % No long form, only description
\newacronym{agv}{AGV}{Autonomous Guided Vehicle}
\newacronym{amr}{AMR}{Autonomous Mobile Robot}
\newacronym{pp}{PP}{Prioritized Planning}
\newacronym{cpAdg}{CP}{Candidate Partitioning}
\newacronym{scpAdg}{SCP}{Sparse Candidate Partitioning}

%% file: common/code_reference_statement.tex
\newcommand{\publicDataStatement}{The evaluation and experimental code used in this study is publicly available at \cite{Dunkel2024EfficientADGConstruction}.}

%% file: common/common_definitions.tex
\newcommand{\GADG}{$G_{\text{ADG}}$ }
\newcommand{\GADGP}{$G_{\text{ADG}}$. }

\newcommand{\MAPFSOLUTION}{\gls{mapf}-Solution }

\newcommand{\NOTE}{\textit{\textbf{Note }}}
\newcommand{\NOTEP}{\textit{\textbf{Note: }}}

\newcommand{\TYPEOne}{\textit{type1 }}
\newcommand{\TYPETwo}{\textit{type2 }}

\newcommand{\InDep}{\rightsquigarrow}

\newcommand{\ADDTypeTwoDep}{\textbf{add} \TYPETwo dependency }

\newcommand{\refAlg}[1]{Algorithm~\ref{#1}}
\newcommand{\refTheorem}[1]{Theorem~\ref{#1}}
\newcommand{\refDefinition}[1]{Definition~\ref{#1}}
\newcommand{\refProposition}[1]{Proposition~\ref{#1}}
\newcommand{\refFigure}[1]{Figure~\ref{#1}}
\newcommand{\refSection}[1]{Section~\ref{#1}}

%% file: src/abstract.tex
\begin{abstract}
\gls{mapf} is critical for coordinating multiple robots in shared environments, yet robust execution of generated plans remains challenging due to operational uncertainties. The \gls{adg} framework offers a way to ensure correct action execution by establishing precedence-based dependencies between wait and move actions retrieved from a \gls{mapf} planning result. The original construction algorithm is not only inefficient, with a quadratic worst-case time complexity it also results in a network with many redundant dependencies between actions. This paper introduces two key improvements to the \gls{adg} framework. First, we prove that wait actions are generally redundant and show that removing them can lead to faster overall plan execution on real robot systems. Second, we propose an optimized \gls{adg} construction algorithm, termed \gls{scpAdg}, which skips unnecessary dependencies and lowers the time complexity to quasi-linear, thereby significantly improving construction speed.

\end{abstract}

%% file: src/introduction.tex
\section{Introduction}

\gls{mapf} is a well-known problem in artificial intelligence, where the goal is to find collision-free paths for multiple agents, typically robots, operating in shared environments \cite{stern_multi-agent_2019}. In real-world applications such as warehouse automation, the complexity of \gls{mapf} increases due to dynamic factors like varying robot speeds, unforeseen obstacles, and continuous task generation. While state-of-the-art \gls{mapf} solvers can generate feasible plans for large robot fleets in relatively short times, robust execution on physical robots—subject to kinematic and dynamic constraints—remains challenging due to operational uncertainties.

In recent years, three key approaches have been developed to address these challenges: (1) Integrate operation uncertainty into the planning process:
This includes methods like  \cite{zhang_priority-based_2024}, \cite{atzmon_robust_2018} or \cite{chen_symmetry_2021}, where the goal is to find $k$-robust plans (where each agents operational delay will not cause conflicts for $k$ time steps). (2) Predictive control of agents during execution: In this category fall works such as \cite{gregoire_locally-optimal_2017} and \cite{wen_distributed_2024}. (3) Post-process a plan into a time independent execution policy: These include works such as \cite{hoenig_multi-agent_2016} and the \glslong{adg} framework, proposed by Hönig et. al \cite{hoenig_persistent_2019}. Due to its simplicity and low communication overhead the \gls{adg}-framework has become a commonly used approach, that has since been adopted in different works such as \cite{varambally_which_2022}, \cite{berndt_feedback_2020}, \cite{berndt_receding_2024} and \cite{liu_multi-agent_2024}.

The \gls{adg} ensures robust execution of \gls{mapf}-plans by establishing dependencies between actions based on their precedence relations. However, the original \gls{adg} construction algorithm exhaustively checks all actions for potential conflicts, leading to a worst-case time complexity of $O(n^2)$, where $n$ is the number of actions. 
Furthermore, this process results in a graph with many redundant dependencies between actions, leading to unnecessary communication and memory overhead.

This paper introduces two key improvements to the \gls{adg} framework. First, we prove that wait actions are redundant and can be removed without affecting execution correctness. Furthermore, we show that removing them can lead to faster overall plan execution (makespan) in scenarios where consecutive motion is expected to be faster than stop-and-go behavior.
Second, we propose an optimized \gls{adg} construction algorithm termed \glslong{scpAdg} that efficiently skips unnecessary dependencies resulting in a significantly faster construction process and a sparser graph without any redundant dependencies.

To the best of our knowledge, no prior work has specifically focused on improving the runtime of \gls{adg} construction itself. While in \cite{berndt_receding_2024}, the authors discuss the conversion of an \gls{adg} into a \gls{seAdg} that does not contain wait actions, they do not directly drop them nor prove that they are unnecessary.

Although solving the \gls{mapf}-problem optimally is NP-complete \cite{yu_structure_2013}, making it the primary computational bottleneck in \gls{mapf}-based systems, our improved \gls{adg} construction process still provides meaningful performance gains, especially in systems utilizing fast planning schemes such as \gls{pp} \cite{erdmann_multiple_1987}.  In lifelong \gls{mapf}, where tasks arrive continuously, \gls{adg} construction and execution must be performed repeatedly. As a result, the performance improvements we introduce become even more significant.

%% file: src/problem_description.tex
\section{Problem Description}

In this section, we describe the \gls{adg} framework and introduce key concepts and definitions used throughout this paper. Further we will discuss the original \gls{adg} construction algorithm and its limitations.

The construction of an \glslong{adg} -- further referred to as $G_{ADG}$ -- begins once a valid \gls{mapf} solution $\Pi$ is found. A solution $\Pi$ is a set of $k$ single-agent plans $\pi_i$ for each agent (robot) $R_i$. A single agent plan $pi_i$ represents a sequence of actions $(a_0, ..., a_n)$ that lead $R_i$ from its start to its goal vertex.

\begin{definition} \label{def:action_definition}
Given a valid \gls{mapf}-Solution $\Pi$, let $A$ be the set of all actions $\{a_0, ..., a_n\}$. An action $a \in A$ is a four-tuple $\langle s, g, t, R \rangle$ with:
\begin{itemize}
  \item \textbf{s}: the start vertex of action $a$,
  \item \textbf{g}: the goal vertex of action $a$,
  \item \textbf{t}: the time step at which action $a$ occurs,
  \item \textbf{R}: the agent (robot) performing action $a$.
\end{itemize}
\end{definition}

\NOTEP While an \gls{adg} consists of action vertices with additional status information \{completed, enqueued, pending\}, this is not relevant for the purpose of this paper. For simplicity, we will use the terms "action vertex" and "action" interchangeably.

\begin{definition} \label{def:no_colliding_actions}
If $\Pi$ is a valid \MAPFSOLUTION, there are no two actions $a$ and $b$ that start or end at the same vertex $v$ and time~step~$t$. More formally: for all $a, b \in A, \; \text{if }  a.t=b.t: \{ a.s \neq b.s  \lor a.g \neq b.g \}$ must hold. (Otherwise, $\Pi$ would contain a collision.)
\end{definition}

Given a valid \MAPFSOLUTION $\Pi$, all its actions $a_i \in A$ are then used to construct the \gls{adg}.

\subsection{Action Dependency Graph Construction}

The graph \GADG is a directed acyclic graph representing the temporal dependencies between actions in a \MAPFSOLUTION $\Pi$. Actions $a \in A$ represent the vertices of the graph, and dependencies between actions are represented as directed edges.
If e.g. there is an edge from action $a$ to $b$, this means that action $b$ depends on action $a$ and can only be executed after $a$ has been completed.

\NOTE that a direct dependency between two actions $a$ and $b$ is denoted as $a \to b$, meaning $b$ depends on $a$. An indirect dependency, such as $a \to c \to b$, is denoted as $a \InDep b$. Since $(a \to b)$ implies $(a \InDep b)$ we will consider a direct dependency $(a \to b)$ as a special case of an indirect dependency $(a \InDep b)$.

\begin{definition} \label{def:type_1_dep_definition}
    Every consecutive action pair $a_i$ and $a_{i+1}$ for the same agent $R_a$ is connected by a \TYPEOne dependency. Formally: for all $a, b \in A$, if $a.R = b.R \land a.t = b.t - 1$ then $a \to b$.
\end{definition}

The original ADG construction algorithm works as follows:
\begin{enumerate}
  \item Connect all consecutive actions of the same shuttle with \TYPEOne dependencies.
  \item Create all \TYPETwo dependencies by exhaustively checking for potential conflicts between actions of different agents. This process is depicted in Algorithm~\ref{alg:og_type2_dep_creation}.
\end{enumerate}

\begin{algorithm}
  \caption{Original ADG-Construction Algorithm}
  \label{alg:og_type2_dep_creation}
  \begin{algorithmic}[1]
    \For {action $a_i \in A$}
      \For {action $a_j \in A$}
        \If {$a_i.R = a_j.R$}
          \State \textbf{continue}
        \EndIf
        \If {$a_j.s = a_i.g$ and $a_j.t \leq a_i.t$}
          \State \ADDTypeTwoDep $a_j \to a_i$
        \EndIf
      \EndFor
    \EndFor
  \end{algorithmic}
\end{algorithm}

Algorithm~\ref{alg:og_type2_dep_creation} exhaustively checks all actions against each other and therefore has a time complexity of $O(n^2)$, where $n$ is the number of actions, in the worst case.

\NOTE that Algorithm~\ref{alg:og_type2_dep_creation} is a simplified version of the original algorithm adapted to our notation. Specifically, the original algorithm discussed in \cite{hoenig_persistent_2019} assumes actions are sorted per agent, and thus avoids iterating over other actions of the same agent, saving some iterations. However, the original algorithm still has a quadratic time-complexity, and therefore performs similarly to our simplified version.

%% file: src/approach.tex
\section{Approach}

The contribution of this paper is two-fold: First, we prove that wait actions are redundant and can be removed from \GADG. Second, we provide a more efficient algorithm for Action Dependency Construction and prove its correctness.

\input{src/approach/wait_actions_redundant}

\input{src/approach/cp_improvement}

\input{src/approach/skipping_redudanent_deps_scp}

%% file: src/approach/wait_actions_redundant.tex
\subsection{Wait Actions are redundant} \label{sec:proof_wait_acitons_are_redundant}

In this section we will prove that wait actions (these are actions where start and goal vertex are the same) are redundant, because they do not contribute any unique dependency to \GADG and can safely be ignored during the construction process.

\begin{proposition} \label{prop:previous_action_definition}
    For every action $b \in A$, if $b.t > 1$ then there exists a \emph{previous} action $a \in A$ with $b.s = a.g \land b.t = a.t + 1 \land b.R = a.R$
\end{proposition}

\begin{definition} \label{def:indirect_implies_reachable}
    For two given actions $a$ and $b$, $b$ is considered \emph{reachable} from $a$ if there exists a direct dependency $a \to b$ or an indirect dependency $a \InDep b$.
\end{definition}

\begin{definition} \label{def:redundant_dependency_definition}
    A dependency $a \to b$ is called \emph{redundant} if removing it does not change the action precedence order. More formally $a \to b$ is \emph{redundant} if $b$ still remains \emph{reachable} from $a$ after dependency $a \to b$ was removed from \GADGP
\end{definition}

To prove that wait actions are generally \emph{redundant} we will show that they do not contribute to any unique dependency in \GADGP First we will show that any outgoing dependency from a wait action is \emph{redundant}.

\begin{theorem} \label{theorem:an_action_dep_on_a_wait_action_also_depends_on_the_next_move_action}
    Suppose agent $R$ has a wait action $w$, and another agent $R'$ has an action $a'$ that depends on $w$ (thus $a'.t \geq w.t \land a'.g = w.s$) then $a'$ also depends on the next move action $m$ after $w$.
\end{theorem}

\begin{proof}
$ $
\begin{itemize}

    \item Since $a'$ depends on $w$, $a'.t > w.t$ must hold because $a'.t = w.t$ would imply that $a'$ and $w$ have the same goal and time step, which contradicts Definition~\ref{def:no_colliding_actions}.
    \item Since wait action $w$ and its subsequent next action $a$ have the same start vertex, $a'$ must also depend on $a$ because $a'.t \geq a.t \land a'.g = a.s$.
    \item Since $a'$ depends on both $w$ and $a$, and $w$ precedes $a$ via a \TYPEOne dependency (Definition~\ref{def:type_1_dep_definition}), the dependency $w \InDep a'$ exists implicitly through $w \to a \to a'$.
    \item If the next action $a$ after $w$ is a wait action this argument can be repeated until the next move action $m$ is reached.
    \item Hence, the direct dependency from $w$ to $a'$ is \emph{redundant}.
\end{itemize}
\end{proof}

Further we will show that any dependency from another action to a wait action is also \emph{redundant}.

\begin{theorem} \label{theorem:wait_actions_do_not_need_dependencies_themselfes}
    Suppose agent $R$ has a wait action $w$, and agent $R'$ has an action $a'$ with $a' \to w.t$, then $w$'s \emph{previous} action $a$ for agent $R$ also depends on $a'$.
\end{theorem}

\begin{proof}
$ $
    \begin{itemize}
        \item Since $w$ depends on $a'$, it follows that $a'.t < w.t$, otherwise $a'$ and $w$ would have the same goal and time step, which contradicts with Definition~\ref{def:no_colliding_actions}.
        \item Since $w.g = a.g$ and $w$ depends on $a'$ then $a$ must also depend on $a'$ because $a'.t \leq a.t $ and $a'.s = a.g$.
        \item Since $a$ depends on $a'$ and $a$ precedes $w$ via a \TYPEOne dependency (Definition~\ref{def:type_1_dep_definition}), the dependency $a' \to w$ exists implicitly through $a' \to a \to w$.
        \item Hence, the direct dependency from $a'$ to $w$ is redundant.
    \end{itemize}
\end{proof}

\textbf{Wait actions are redundant in the ADG framework}
\\
Given Theorem~\ref{theorem:an_action_dep_on_a_wait_action_also_depends_on_the_next_move_action}, any \TYPETwo dependency from a wait action $w$ to another agent action $b$ is redundant and given Theorem~\ref{theorem:wait_actions_do_not_need_dependencies_themselfes}, any \TYPETwo dependency from another agent action $b$ to a wait action $w$ is redundant.
Thus, wait actions do not add any unique dependencies and can be safely removed from \GADG without altering the overall action execution flow. Not only are wait actions redundant, removing them improves the \gls{adg} framework in the following ways:
\begin{enumerate}
    \item \textbf{Improved construction efficiency}: Fewer actions and dependencies need to be added, accelerating the overall process.
    \item \textbf{Sparser graph}: This reduces communication overhead during execution, making the system more efficient.
    \item \textbf{Increased system throughput}: If robot $R_a$ is waiting for robot $R_b$ to finish its action, $R_a$ can begin its next task immediately once $R_b$ completes, rather than waiting for the duration of a scheduled wait action. This is particularly beneficial when $R_b$ finishes earlier than planned, allowing $R_a$ to start its next action sooner than if wait actions were used.
\end{enumerate}

%% file: src/approach/cp_improvement.tex
\subsection{Improved ADG-Construction via Candidate Partitioning}
After proving that wait actions are redundant and can be safely ignored in the \gls{adg} construction process, we now turn our focus to further optimizing the construction algorithm itself. Instead of exhaustively checking all actions against each other, for each action $a_i \in A$, only looking at other actions $a_j$ where $a_j.s = a_i.g$ is sufficient.
These actions are further denoted as candidate actions $C$.

\begin{definition} \label{def:candidates_are_all_that_matters}
    For any action $a \in A$ its candidate list $C [c_0, ..., c_n]$ represents all actions where a \TYPETwo dependency $c_i \to a$ is possible.
\end{definition}

Based on this insight, it is easy to create an improved \gls{adg}-construction algorithm. This works as follows:

\begin{enumerate}
    \item Remove all wait actions from $A$.
    \item Connect all consecutive actions of the same shuttle with \TYPEOne dependencies.
    \item Create a \textbf{candidate action lookup} $(S)$:
    $S$ is a mapping from a vertex $v$ to a list of candidate actions $C$ that have $v$ as their start vertex. \NOTE that any efficient key-value lookup data structure can be used for this purpose.
    \item For each action $a_i \in A$ only consider the candidate actions $c_i$ from $S[a_i.g]$ and add a dependency $c_i \to a_i$ if $c_i.R \neq a_i.R \land c_i.t < a_i.t$ holds.
\end{enumerate}
This process is depicted in Algorithm~\ref{alg:adg_construction_via_candidate_action_lookup}.

Partitioning actions based on their start vertices significantly reduces the number of comparisons needed to create \TYPETwo dependencies in practice. While the worst-case asymptotic complexity remains quadratic at $O(n^2)$, the average runtime is reduced to $O(n^2 / x)$, where $x$ is the number of distinct start vertices in the graph. This reduction occurs because all actions in a \gls{mapf} solution are typically evenly spread across the grid, resulting in a smaller number of candidate actions for each action.

\begin{algorithm}
  \caption{Candidate Partitioning (CP) ADG-Construction}
  \label{alg:adg_construction_via_candidate_action_lookup}
    \begin{algorithmic}[1]
    \State $S \gets \{\}$ 
    \For{action $a_i \in A$}
        \State $S[a_i.s] \gets S[a_i.s] + a_i$  
    \EndFor
    \For{action $a_i \in A$}
        \State $C \gets S[a_i.g]$  
        \For{action $c_i \in C$}
            \If{$c_i.R \neq a_i.R$ and $c_i.t \leq a_i.t$}
                \State \ADDTypeTwoDep $c_i \rightarrow a_i$
            \EndIf
        \EndFor
    \EndFor
    \end{algorithmic}
\end{algorithm}

%% file: src/approach/skipping_redudanent_deps_scp.tex
\subsection{Skipping Redundant Dependencies with Sparse Candidate Parititioning (SCP)} \label{sec:scp}

Having established an improved \gls{adg} construction algorithm based on candidate partitioning, we now turn to further optimizing the process by skipping redundant dependencies. \gls{scpAdg} leverages the insight that many dependencies between an action $a$ and its candidate actions $c_i \in C$ are redundant. Specifically, it is sufficient to only add a dependency from the latest (different robot) candidate action, as defined in \refTheorem{theorem:only_the_overall_last_action_is_relevant}.

\begin{theorem} \label{theorem:only_the_overall_last_action_is_relevant}
    For a given action $a \in A$ and its candidate list $C [c_0, .., c_n]$, only adding a dependency $c_k \to a$, with $c_k$ being the latest action in $C$ with $c_k.t \leq a_i.t \land c_k.R \neq a_i.R$, still maintains an indirect dependency $c_i \InDep a$ for all $c_i \in C$, and thus is sufficient in maintaining execution correctness in the \gls{adg} framework.
\end{theorem}

\subsubsection{SCP Construction Algorithm}

Based on \refTheorem{theorem:only_the_overall_last_action_is_relevant} we can create a new ADG-Construction algorithm that skips redundant dependencies. This algorithm is depicted in 
\refAlg{alg:adg_construction_without_redundant_type2_deps}.

\begin{algorithm}
\caption{ ADG-Construction}
\label{alg:adg_construction_without_redundant_type2_deps}
    \begin{algorithmic}[1]
    \State $S \gets \{\}$
    \For{action $a_i \in A$}
        \State $S[a_i.s] \gets S[a_i.s] + a_i$ 
     \EndFor
     \For{$(s, C) \in S$}
        \State $C \gets $ sort($C$) by $t$ in ascending order.
     \EndFor
     \For{action $a_i \in A$}
        \State $C \gets S[a_i.g]$  
        \State $k \gets $ bisect\_right($C, a_i.t$) - 1
        \If{$k < 0 \lor c_k.R = a_i.R$}
            \State $k \gets k - 1$
            \State \textbf{continue}
        \EndIf
        \State $c_k \gets C[k]$
        \State \ADDTypeTwoDep $c_k \to a_i$
    \EndFor
    \end{algorithmic}
\end{algorithm}

In (1-3) a candidate action lookup $S$ that contains a mapping from a vertex $v$ to a list of candidate actions $C$ that have $v$ as their start vertex is created. In (4-5) every candidate list $C_i \in S$ is sorted by time in ascending order. In (6-13) \TYPETwo dependencies are created for every action $a_i \in A$. This process works similarly to \refAlg{alg:adg_construction_via_candidate_action_lookup}, but instead of iterating over all candidate actions $c_i \in C$, a binary search for the latest candidate action $c_k$ with $c_k.t \leq a.t$ and $c_k.R \neq a.R$ is done (8-12). If a $c_k$ is found, only a dependency $c_k \to a_i$ is created and no other (13).

\subsubsection{Proving \refTheorem{theorem:only_the_overall_last_action_is_relevant}}

\begin{corollary}
    As defined in \refDefinition{def:indirect_implies_reachable}, \refTheorem{theorem:only_the_overall_last_action_is_relevant} implies that for each action $a_i \in A$, $a_i$ is \emph{reachable} from all of its candidate actions $c_i \in C$ with ${c_i.t \leq a_i.t}$.
\end{corollary}

Therefore, to prove \refTheorem{theorem:only_the_overall_last_action_is_relevant} we will show that every candidate action $c_i \in C$ is \emph{reachable} from $a$. In other words: we will prove that for every $c_i \in C$ there is an indirect dependency $c_i \InDep a$. To prove this it is sufficient to show that there is a dependency between every two consecutive candidate actions $c_i$ and $c_{i+1} \in C$.

\begin{proof} 
$ $
\begin{itemize}
    \item Given by \refDefinition{def:no_colliding_actions}, it is obvious that no two consecutive candidate actions $c_i$ and $c_{i+1}$ can have the same time step and thus: $c_i.t < c_{i+1}.t$ 
    \item Given by \refProposition{prop:previous_action_definition}, for $c_{i+1}$ there is a \emph{previous} action $b_{i+1}$ with $b_{i+1}.t = c_{i+1}.t - 1$ and thus: $c_i.t \leq b_{i+1}.t$. 
    \item To show that $c_i \InDep c_{i+1}$ it is sufficient to show that $c_i \to b_{i+1}$ holds.
    \item When \gls{scpAdg} eventually iterates over $b_{i+1}$ and a dependency is added between $b_{i+1}$ and its last candidate action $c_k$, it is therefore necessary that $c_k = c_i$ to ensure that $c_i \to b_{i+1}$ holds.
    \item Since $b_{i+1}.g = a.g$ they also share the same candidate set containing $c_i$. 
    \item Since no two actions in $A$ can share the same start and time step, it is guaranteed that $c_i$ is the latest candidate action in $C$ with $c_i.t \leq b_{i+1}.t$. Therefore, applying the algorithm to action $b_{i+1}$ creates a dependency $c_i \to b_{i+1}$.
    \item Thus, applying the algorithm to every action in $A$ will create dependencies between every two consecutive candidate actions $c_i$ and $c_{i+1} \in C$.
    
\end{itemize}
\end{proof}

\subsubsection{Runtime Analysis of \gls{scpAdg}}

The runtime complexity of the \gls{scpAdg} algorithm is derived from two key operations:
\begin{enumerate}
    \item Sorting of Candidate Lists: First a candidate list $C$ is constructed for every distinct action start vertex $x$. In practice the number of actions in each candidate list is limited. Asymptotically this step has a runtime of $O(n  \log(n))$ where $n$ is the total number of actions.
    \item Binary Search and Dependency Creation: For each action $a_i$, our algorithm performs a binary search on the sorted candidate list to find the latest relevant candidate action. This search operation is $O(\log(n))$ for each action. Given that this is done for every action in $A$, the overall complexity for this step is $O(n  \log(n))$.

    Overall, the total runtime complexity of the \gls{scpAdg} algorithm is $O(n  \log(n))$, making it significantly more efficient than the original exhaustive construction algorithm which has a complexity of $O(n^2)$.
\end{enumerate}

%% file: src/experimental_evaluation.tex
\section{Experimental Evaluation}

In this section, we evaluate two key aspects of our contributions: first, we compare the proposed \gls{adg} construction algorithms, and second, the impact of removing wait actions on overall execution performance. The following experiments compare the efficiency of our newly developed \gls{adg} construction algorithms and demonstrate how omitting wait actions reduces the system's makespan by better taking advantage of faster-then planned action execution.

To evaluate our proposed improvements, we selected the \gls{mapf} benchmark introduced by Tan et al., which includes six representative maps from the official \gls{mapf} benchmark suite \cite{stern_mapf_benchmark}. Each map contains 125 unique scenarios, equally distributed across five different agent counts. Specifically, we utilize the authors' pre-computed \gls{mapf} solutions. For further details we refer the reader to \cite{tan_benchmarking_2024}.

% \anonymousDataStatement
\publicDataStatement

\subsection{\gls{adg}-Construction Algorithm Comparison}

To evaluate our newly developed \gls{adg}-Construction procedures, specifically \glslong{cpAdg} 
(\refAlg{alg:adg_construction_via_candidate_action_lookup}) and \glslong{scpAdg} (\refAlg{alg:adg_construction_without_redundant_type2_deps})
we conducted a series of experiments.
\NOTE that we implemented all the construction algorithms in C++ to ensure representative runtime measurements.

\subsubsection{\gls{adg}-Construction Runtime Comparison}
\begin{figure}[h]
  \centering
  \includegraphics[width=1\linewidth]{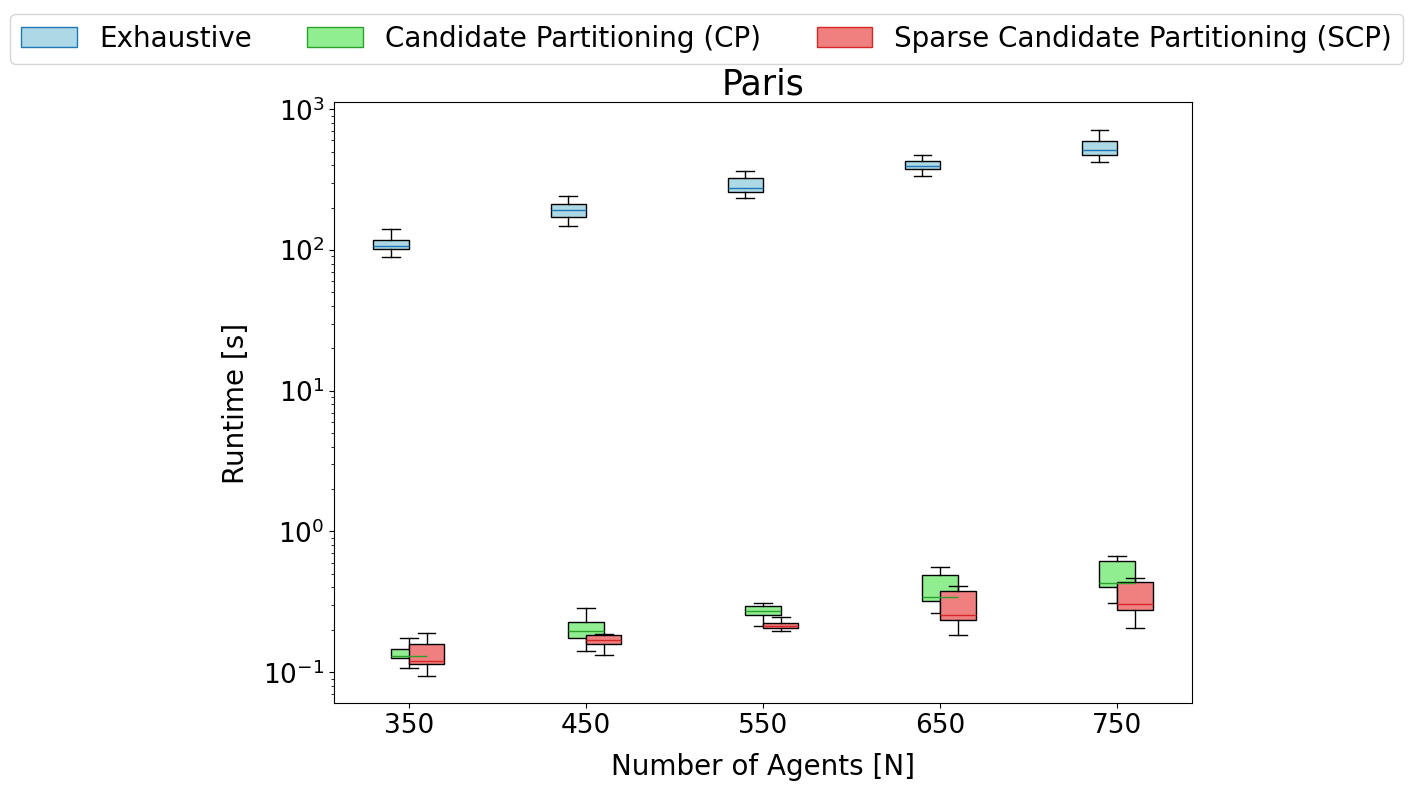}
  \caption{Runtime comparison of \gls{adg} construction algorithms on the Paris map, comparing the exhaustive method with \gls{cpAdg} and \gls{scpAdg}. The exhaustive method shows an orders of magnitudes worse runtime than both \gls{cpAdg} and \gls{scpAdg}.}
  \label{fig:exhaustive_vs_others}
  \Description{The runtime comparison between three \gls{adg} construction algorithms (Exhaustive, \glslong{cpAdg}, and  \glslong{scpAdg}) on the Paris map. The x-axis represents the number of agents, and the y-axis shows runtime in seconds (logarithmic scale). The Exhaustive method exhibits an orders of magnitude worse runtime behaviour, while \gls{cpAdg} and \gls{scpAdg} maintain significantly lower runtime.}
\end{figure}

First we compare the construction runtime of both \gls{cpAdg} and \gls{scpAdg} against the exhaustive approach (\refAlg{alg:og_type2_dep_creation}).
As shown in \refFigure{fig:exhaustive_vs_others}, the exhaustive method suffers from a quadratic increase in runtime as the number of agents grows. When scaling to larger agent sets, both \gls{cpAdg} and \gls{scpAdg} show far superior performance.

\begin{figure}[h]
  \centering
  \includegraphics[width=1\linewidth]{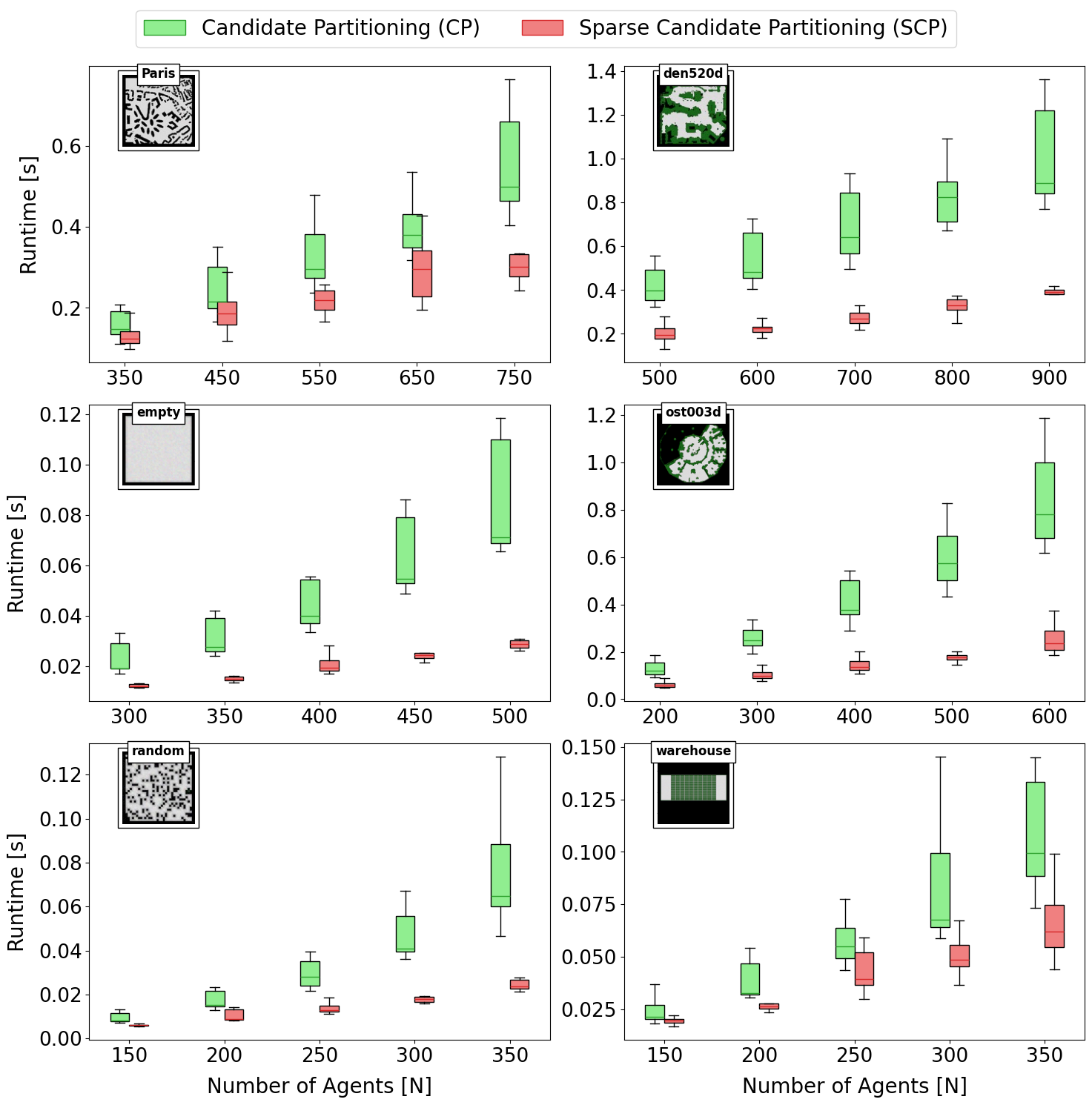}
  \caption{Comparison of \gls{adg}-Construction runtime between \gls{cpAdg} and \gls{scpAdg}. \gls{scpAdg} demonstrates improved overall runtime performance for all maps in the benchmark.}
  \label{fig:cp_vs_scp_box_plot}
  \Description{The runtime comparison between the two fast \gls{adg} construction algorithms (\glslong{cpAdg}, and  \glslong{scpAdg}) on across all maps (Paris, den520d, empty, ost003d, random, warehouse) and varying numbers of agents. The x-axis represents the number of agents, while the y-axis shows runtime in seconds. \gls{scpAdg} demonstrates lower run times compared to \gls{cpAdg}, particularly for larger agent numbers, making it better scaling solution.}
\end{figure}

Given that the exhaustive method scales significantly worse for larger numbers of agents, we provide a detailed comparison on the full benchmark only between \gls{cpAdg} and \gls{scpAdg}. \refFigure{fig:cp_vs_scp_box_plot} presents these runtime comparisons. While \gls{cpAdg} is already very fast, it exhibits a clear upward trend as the number of agents increases. \gls{scpAdg} on the other hand, has an asymptotic runtime of $O(n \log(n))$ which is reflected by the results. In practice, although \gls{scpAdg} is faster than \gls{cpAdg}, both algorithms generally maintain sub-second runtime. Both perform well across our benchmark for agent counts between 150 and 900, though with a further increase in the number of agents, the differences in performance become more pronounced.

\subsubsection{Influence of \gls{adg}-Construction Algorithm on created \TYPETwo Dependencies.}

\begin{figure}[h]
  \centering
  \includegraphics[width=1\linewidth]{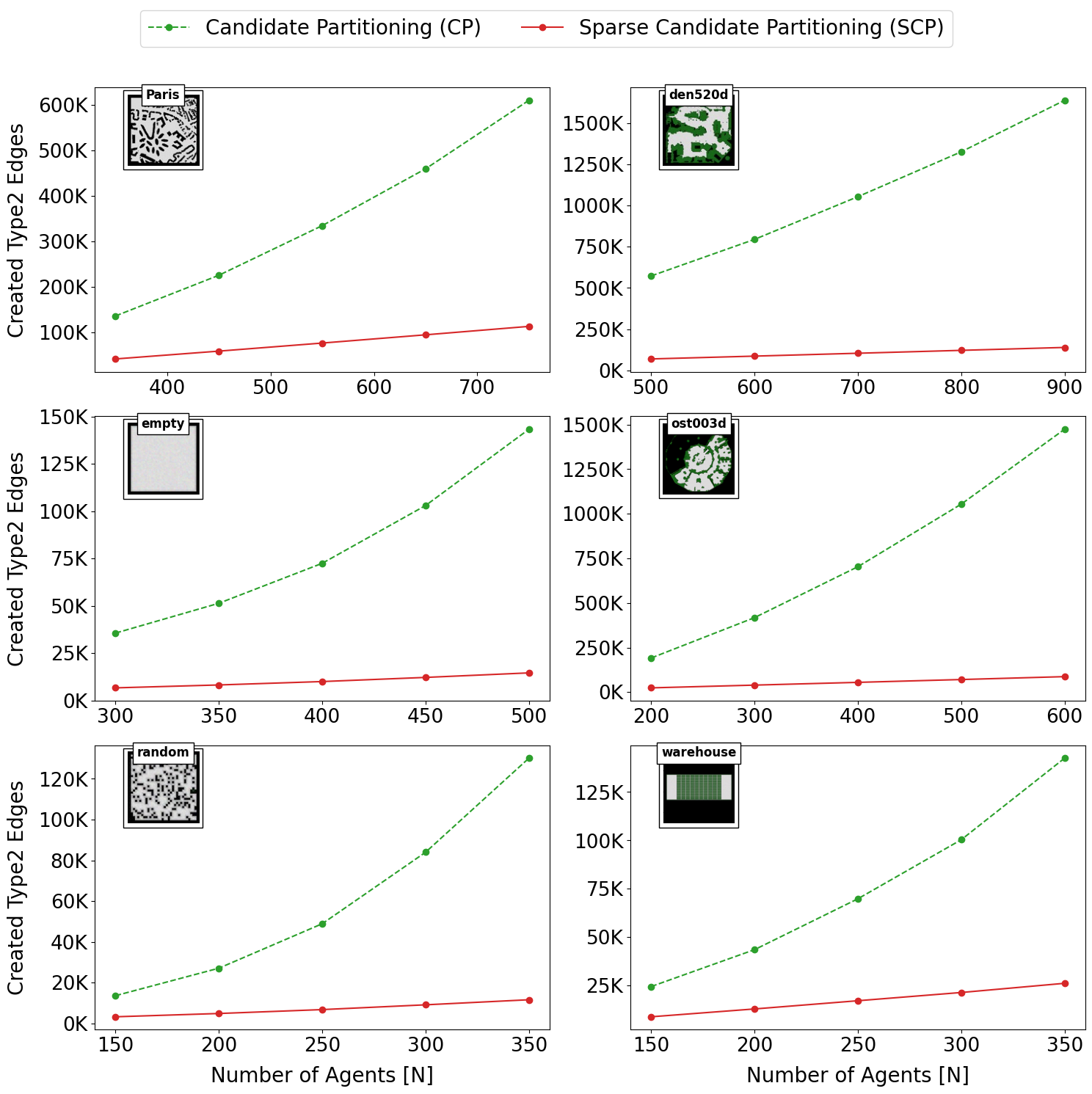}
  \caption{Comparing \gls{cpAdg} and \gls{scpAdg} on the number of created \TYPETwo dependencies across all maps in the benchmark. \gls{scpAdg} generates significantly fewer \TYPETwo dependencies then \gls{cpAdg}}
  \label{fig:created_type2_deps}
  \Description{This figure illustrates the number of Type 2 dependencies created by the \glslong{cpAdg} and \glslong{scpAdg} algorithms across six different maps (Paris, den520d, empty, ost003d, random, and warehouse) and various agent counts. The x-axis represents the number of agents, while the y-axis indicates the total number of Type 2 dependencies created. The number of \TYPETwo dependencies created with \gls{cpAdg} exhibits a clear upward trend as the number of agents increases. \gls{scpAdg} on the other hand only shows a linear increase in dependencies, remaining significantly less overall.}
\end{figure}

As described in \refSection{sec:scp}, constructing the \gls{adg} with \gls{scpAdg} leads to a graph where every action has at most one \TYPETwo dependency. This is a substantial reduction compared to both the original construction algorithm and \gls{cpAdg}. \refFigure{fig:created_type2_deps} illustrates this comparison.
\gls{cpAdg} shows a clear upward trend in the number of created \TYPETwo dependencies as the number of agents increases. This indicates that the number of created \TYPETwo dependencies is currently a limiting factor in scaling the \gls{adg} framework to a very large number of agents. 
\gls{scpAdg} effectively addresses this challenge.

\subsection{Influence of Wait Actions on Execution Performance}

In Section \ref{sec:proof_wait_acitons_are_redundant}, we proved that wait actions are redundant and can be removed from the \gls{adg} without altering the overall action execution flow. In addition to a more streamlined graph, a slightly faster construction process, and reduced agent communication overhead, the removal of wait actions also influences overall execution performance.
To demonstrate this impact, we conducted a series of experiments. Given our chosen benchmark, we selected two maps (Paris, warehouse) to evaluate the \gls{adg} execution duration on all their given pre-computed \gls{mapf}-solutions. These solutions were used to construct \GADGP \NOTE that some pre-computed solutions from the benchmark led to cycles within the \gls{adg} due to the planner not accounting for \gls{adg} constraints and where thus excluded. Then during execution, each agent communicates its actions within the \gls{adg} framework until they eventually run out of actions.

In real-world scenarios, the time required for an agent (robot) to execute an action (e.g. moving between two adjacent cells) is not constant but depends on the robot's internal kinematic constraints. One key advantage of the \gls{adg} framework is that it allows robots to perform continuous motion across multiple consecutive move actions, thereby avoiding unnecessary stop-and-go behavior. When a robot can move directly through multiple locations, it reaches its goal more quickly.
To model plan execution within the \gls{adg} framework, we developed a discrete-event simulation where agents perform actions and, upon completion, advance the simulation time. If the action was a wait action or a single move action, the time advances by 1 second. However, if there is another move action immediately following in the shuttle's queue, the current move action will only take 0.8 seconds, reflecting the faster consecutive movement compared to the typical stop-and-go behavior.

\begin{figure}[h]
  \centering
  \includegraphics[width=1\linewidth]{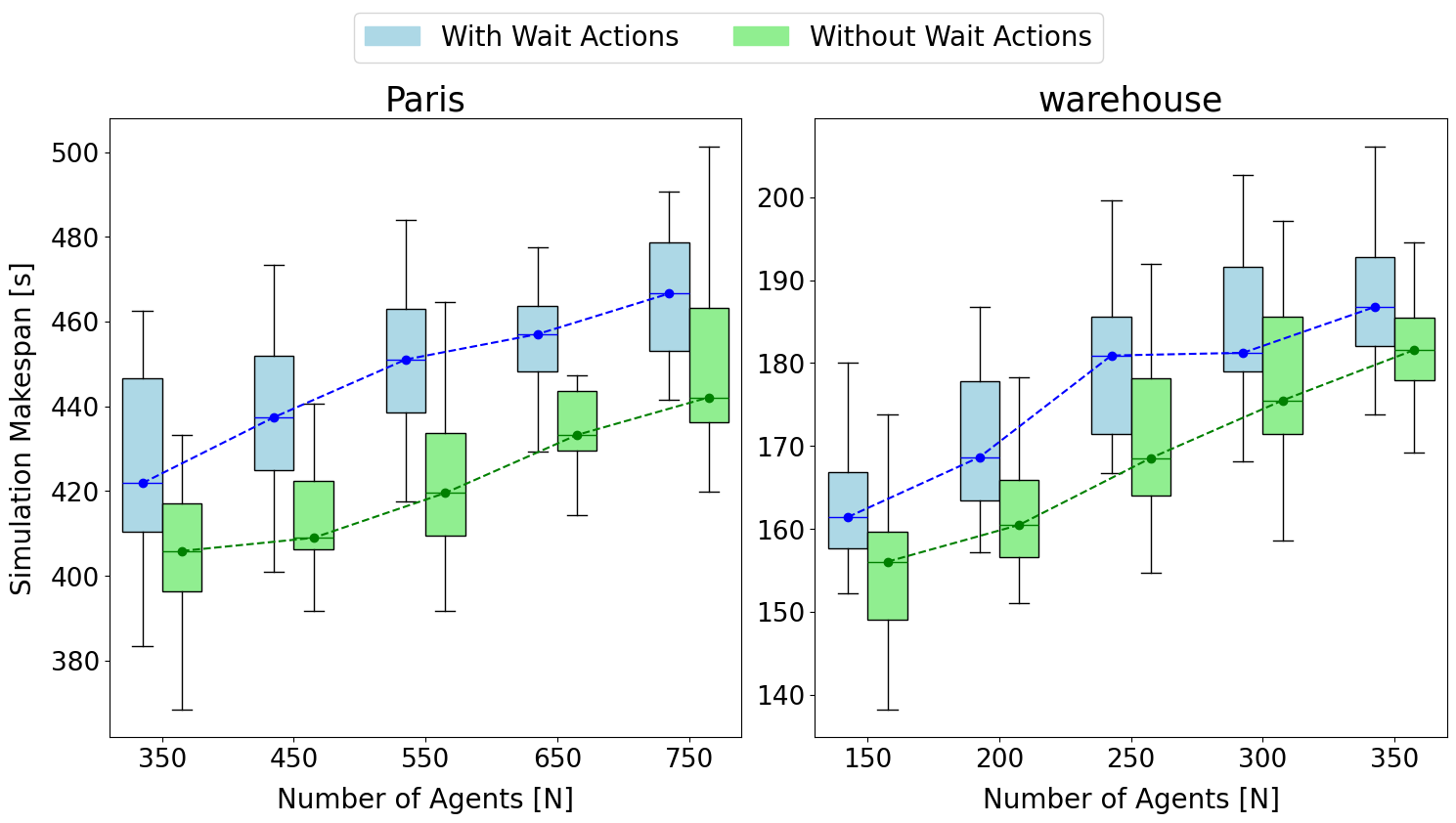}
\caption{Comparison of makespan for simulations with (blue) and without (green) wait actions across varying numbers of agents. Simulations without wait actions generally finish a few seconds earlier, due to allowing agents to better take advantage of consecutive movement.}
  \label{fig:wait_compare_plot}
  \Description{This figure compares the makespan of simulations with and without wait actions across varying agent counts. The x-axis represents the number of agents, while the y-axis shows the total makespan in seconds. Simulations without wait actions (green) consistently show a lower makespan than those with wait actions (blue), illustrating the performance improvement when agents are not constrained by predetermined timing.}
\end{figure}

\refFigure{fig:wait_compare_plot} shows the results of these experiments. The makespan when wait actions are included (blue) is generally higher than when they are omitted (green). This difference in makespan likely results from situations where agents complete their actions faster than planned due to consecutive motion. In such cases, subsequent agents can begin their actions sooner, leading to a cascading effect that reduces the overall makespan. In contrast, when wait actions are used, agents are restricted to their predetermined timing, even when a preceding agent finishes early, which limits these potential optimizations.

%% file: src/conclusion.tex
\balance

\section{Conclusion}

In this paper, we introduced two key improvements to the \glslong{adg} framework: the removal of redundant wait actions and an optimized construction algorithm. Our experiments demonstrate that omitting wait actions consistently improves overall makespan by taking advantage of faster-than-planned consecutive motion. 
Further, the proposed \glslong{scpAdg} algorithm significantly reduces the runtime of the \gls{adg} construction process, with a practical and theoretical complexity of $O(n \log(n))$.
\gls{scpAdg} also generates a graph with at most one \TYPETwo dependency per action, resulting in reduced memory overhead and minimized communication requirements. We believe these improvements not only enhance the scalability of the \glslong{adg} framework but also make it a more practical solution for adoption in systems handling larger agent sets in the future.